%% file: main.tex
\documentclass[11pt]{article}

\usepackage[final]{acl}

\usepackage{times}
\usepackage{latexsym}
\usepackage{amsmath}
\usepackage{amssymb}

\usepackage[T1]{fontenc}

\usepackage[utf8]{inputenc}

\usepackage{microtype}

\usepackage{inconsolata}

\usepackage{graphicx}
\usepackage{booktabs}
\usepackage{multirow}
\usepackage{stfloats}
\usepackage{placeins}

\newcommand{\ourmethod}{HoliTok}

\title{HoliTok:A Continuous Holistic Tokenization with Robust Dual Capabilities\\ of Speech Generation and Understanding}

\author{
 \textbf{Bohan Li\textsuperscript{1}},
 \textbf{Shi Lian\textsuperscript{2}},
 \textbf{Hankun Wang\textsuperscript{1}},
 \textbf{Yiwei Guo\textsuperscript{1}},
 \textbf{Yu Xi\textsuperscript{1}},
 \textbf{Zhihan Li\textsuperscript{1}},
\\
 \textbf{Da Zheng\textsuperscript{2}},
 \textbf{Colin Zhang\textsuperscript{2}},
 \textbf{Kai Yu \textsuperscript{1*}},
\\
 \textsuperscript{1}X-LANCE Lab, School of Computer Science, Shanghai Jiao Tong University, China
\\
 \textsuperscript{2}Xiaohongshu Inc, China
\\
 \small{
   {\{everlastingnight, kai.yu\}@{sjtu.edu.cn}}
 }
}

\begin{document}
\maketitle

\renewcommand{\thefootnote}{\fnsymbol{footnote}}
\footnotetext[1]{is the corresponding author.}

\begin{abstract}
Unified speech foundation models require a holistic tokenization space that is both learnable by language models and decodable into high-quality waveforms. Existing speech tokenizers, however, often fail to satisfy these requirements simultaneously, leading to increased architectural complexity and more involved training designs. We propose \ourmethod{}, a continuous \textbf{Holi}stic speech \textbf{Tok}enization model designed for unified generation-understanding modeling. \ourmethod{} encodes 48~kHz speech into a compact 25~Hz sequence of 128-dimensional latents. It is trained with a progressive strategy that jointly preserves signal-level fidelity, incorporates semantic information, and maintains strong latent learnability. Based on this tokenization, we build a unified AR+DiT model for speech synthesis and recognition, where the same latent sequence supports both generation-specific and unified generation-understanding tasks. Experiments show that \ourmethod{} achieves competitive reconstruction fidelity, improves generative learnability for high-quality and controllable synthesis, and, among the evaluated representations, is the \textit{only} one that operates robustly in our unified generation-understanding architecture without additional optimization tricks. These results suggest that \ourmethod{} serves as an effective speech tokenizer and a foundational representation interface for unified spoken language modeling. The code is available at: \url{https://github.com/bovod-sjtu/HoliTok}.


\end{abstract}

\input{intro}

\input{related_works}

\input{method}

\input{experiment/exp}

\vspace{-5pt}
\section{Conclusion}
\vspace{-5pt}

We present \ourmethod{}, a holistic speech tokenizer for both generation-oriented and unified generation--understanding tasks. Through progressive training, \ourmethod{} combines compact high-fidelity reconstruction, sequence-aware variational regularization, and downstream-aware semantic enrichment, yielding a tokenization that remains detokenizable, learnable, and informative. Experiments on reconstruction, zero-shot and controllable TTS, and unified ASR--TTS modeling demonstrate that \ourmethod{} serves as an effective interface for speech compression, diverse speech synthesis, and unified spoken-language modeling. Comprehensive analyses further show that \ourmethod{} achieves robust performance without relying on complex architectural modifications or incremental training mechanisms.

\section*{Limitations}
This work has two main limitations, both of which point to natural directions for future research. First, our current study focuses on speech-centered generation and understanding. Although \ourmethod{} is designed as a continuous audio representation, our experiments mainly cover speech reconstruction, text-to-speech synthesis, and automatic speech recognition. We have not yet systematically evaluated whether the same latent space can generalize to broader audio domains such as environmental sound and music. These domains may require different temporal abstractions, perceptual objectives, and semantic supervision signals. Extending \ourmethod{} from speech to general audio and music modeling is therefore an important direction for future work. Second, our downstream evaluation is built on a unified AR+DiT architecture. This setting directly tests whether the learned representation can serve as a shared interface for both speech generation and understanding, but it does not exhaust all possible unified modeling paradigms. In particular, we have not explored pure DiT-based or fully non-autoregressive architectures for unified generation-understanding modeling. Future work can study how \ourmethod{} interacts with different backbone designs, and whether the proposed representation remains robust across alternative generative and understanding architectures. Potential risks, artifact documentation, computational experiment details, and AI-assistant use are further discussed in Appendices~\ref{app:setting_details} and~\ref{app:ai_assistants}.


\bibliography{custom}

\input{appendix}

\end{document}

%% file: intro.tex
\section{Introduction}

Recent progress in multimodal foundation models is moving toward unified understanding and generation~\cite{glm4voice, kimiaudio, ge2025seedxmultimodalmodelsunified, fan2025unifiedautoregressivevisualgeneration,xie2025showo,xie2026show}. Rather than treating downstream tasks separately, emerging systems seek to build all-in-one architectures that can understand, reason over, and generate within a shared parameter space. In the speech domain, this direction places a stronger requirement on the tokenizer: speech should be represented in a continuous space that is simultaneously decodable, learnable, and informative, so that it can serve as the interface for unified generation-understanding modeling. However, such a holistic continuous speech tokenizer remains underdeveloped. In its absence, downstream models must compensate through incremental architectural designs, such as task-specific encoders, multiple token streams, or decoupled modules. Consequently, the burden of unification is shifted from the representation itself to increasingly complex model design~\cite{xu2025qwen25omnitechnicalreport, xu2025qwen3omnitechnicalreport,yan2025minguniaudiospeechllmjoint}.

Conventional acoustic front-end features, such as mel spectrograms, Fbank features, and MFCCs~\cite{mfcc}, retain local signal structure, but they produce dense frame-level sequences that are redundant and difficult to model for downstream understanding and generation. In contrast, self-supervised speech representations~\cite{wav2vec2, hubert, wavlm} expose richer semantic information, but they are not naturally decodable into high-fidelity waveforms and often present a challenging target for generative modeling. Thus, existing representations typically satisfy only part of the requirements for unified continuous speech modeling, leaving a gap between semantic abstraction, acoustic fidelity, and model learnability.

Current speech tokenizers address this challenge only partially. Discrete codec-based tokenizers~\cite{défossez2022highfidelityneuralaudio, ji2025wavtokenizer, du2024cosyvoice, yiwei_survey} compress speech into language-model-friendly symbols, but quantization and multi-codebook designs may introduce information loss and additional modeling complexity. Continuous tokenizers~\cite{li-etal-2025-continuous, niu2025semanticvae, cheng2026distillationlossfunctionsspeech} avoid quantization and are favorable for generation, yet many are optimized mainly for reconstruction or synthesis rather than as a shared tokenization space for unified generation-understanding models. Existing ``unified'' representations~\cite{dinkel2026dashengtokenizerlayerunifiedaudio, yang2026wavcubeunifyingspeechrepresentation} are also often evaluated in task-specific systems separately, leaving the consistency of the shared modeling space unclear.

Recent AR+DiT architectures offer a simple downstream framework for unified continuous speech generation and understanding. For example, Ming-UniAudio~\cite{yan2025minguniaudiospeechllmjoint} proposes MingTok-Audio to connect a compact variational autoencoder (VAE) latent with richer semantic features via an additional semantic module. While this improves tokenizer usability, the low-level latent remains fixed as higher-level semantics are introduced, resulting in an inconsistent modeling space and limited generative capacity.

In this work, we propose \textbf{\ourmethod{}}, a \textbf{Holi}stic \textbf{Tok}enization model for unified continuous speech generation and understanding. \ourmethod{} encodes 48~kHz speech into a compact 25~Hz sequence of 128-dimensional continuous latents. Its training follows a progressive recipe that gradually shapes a learnable and semantically informative latent space. We first train an autoencoder to ground the representation in faithful waveform reconstruction. We then introduce a sequence-aware variational bottleneck to regularize the latent distribution, making the sequence smoother and easier to predict while preserving signal-level fidelity. Finally, we strengthen variational regularization and refine the latent space through high-level feature distillation and audio-language supervision, enabling the resulting tokenization to retain information useful for spoken language understanding while remaining highly learnable for diverse speech synthesis tasks.

We build a unified generation-understanding model based on an AR+DiT architecture to evaluate whether a continuous speech tokenizer can serve as a unified modeling interface. The latent sequence is first encoded into patch embeddings for autoregressive modeling by the LLM. For generation, the LLM predicts semantic hidden states, which condition a DiT-based flow-matching head to predict the next latent patch. For understanding, the LLM predicts the next text token through an LM head. This evaluation is intentionally downstream-aware: beyond measuring reconstruction quality, it examines whether the tokenizer facilitates unified AR+DiT modeling.

We evaluate \ourmethod{} from three complementary perspectives: reconstruction, speech synthesis, and unified generation-understanding modeling. Empirically, \ourmethod{} achieves competitive reconstruction fidelity with a highly compact latent sequence, while supporting high-quality, diverse, and controllable TTS. In unified spoken language modeling, instantiated with ASR and TTS, \ourmethod{}-Base already provides a more modeling-friendly continuous latent space than existing alternatives. \ourmethod{}-Unite further improves both synthesis and recognition by incorporating the causal semantic encoder trained in the final stage, demonstrating substantially better usability than the baselines. These results show that \ourmethod{} is not only an effective speech tokenizer, but also a principled representation interface that bridges the modeling-space gap between unified continuous speech understanding and generation.

%% file: related_works.tex
\section{Related Work}

\paragraph{Audio representation for unified generation and understanding.}
Audio tokenization for unified modeling has been studied through both discrete and continuous representations. Discrete codecs and speech tokenizers~\cite{défossez2022highfidelityneuralaudio, ji2025wavtokenizer, du2024cosyvoice}, provide compact language-model-friendly units.  Continuous tokenizers avoid quantization and have been explored in ~\cite{li-etal-2025-continuous,niu2025semanticvae, dinkel2026dashengtokenizerlayerunifiedaudio, yang2026wavcubeunifyingspeechrepresentation} for speech synthesis or unified audio modeling. These works improve different aspects of acoustic fidelity, semantic accessibility, and downstream usability. Compared with these works, \ourmethod{} emphasizes holistic evaluation of the tokenization space within a single unified generation--understanding model, directly testing whether the same continuous representation is modelable as a shared interface for both speech generation and understanding. 

\paragraph{Unified generation-understanding architecture with continuous tokens.}
Continuous-token architectures have recently emerged for unified generation and understanding. In vision, recent works~\cite{fan2025unifiedautoregressivevisualgeneration, xie2025showo,xie2026show} perform autoregressive generation and understanding with continuous visual tokens. Similar in audio, DiTAR~\cite{jia2025ditar} uses an autoregressive backbone with a DiT-based flow-matching head for continuous speech patches, and Ming-UniAudio~\cite{yan2025minguniaudiospeechllmjoint} extends this idea to unified speech understanding, generation, and editing. Our contribution lies in the tokenizer rather than the AR+DiT backbone. Unlike MingTok-Audio, which uses a semantic representation for understanding and an acoustic latent as the generation target, \ourmethod{} progressively shapes its VAE latent space for reconstruction, generation, and understanding. We therefore use the shared AR+DiT setting as a controlled evaluation protocol.

%% file: method.tex
\section{Methodology}

\begin{figure*}[t]
\centering
\includegraphics[width=\textwidth]{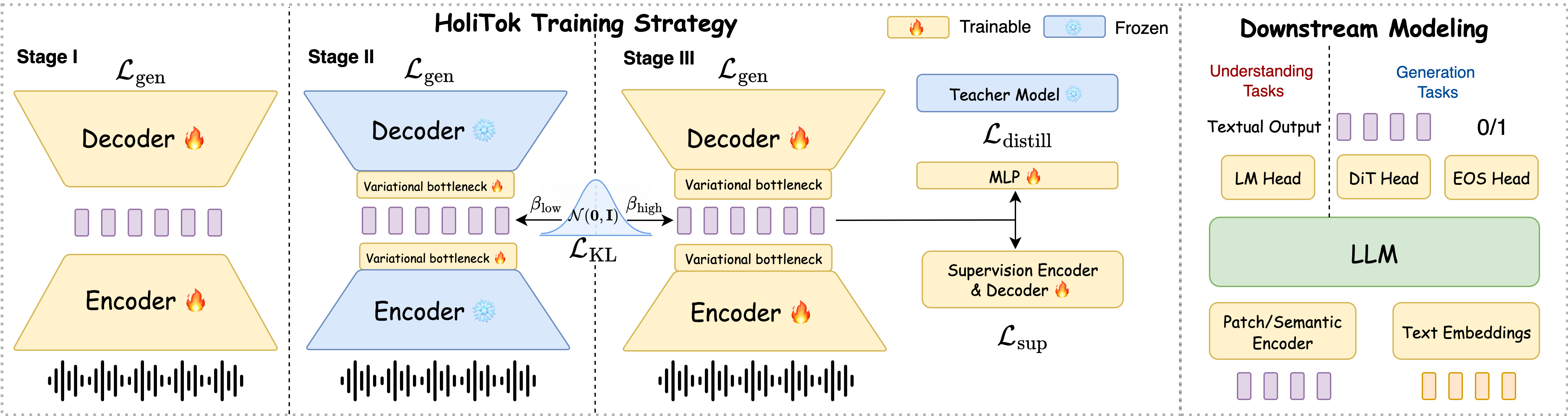}
\caption{An Overview. Left side is the three-stage training strategy of \ourmethod{}; Right side is our downstream architecture for unified generation-understanding tasks.}
\label{fig:overview}
\vspace{-9pt}
\end{figure*}

\vspace{-4.5pt}
\subsection{Main Architecture}
\label{sec:architecture}

\ourmethod{} is a speech tokenizer built on a low-latency variational autoencoder backbone. We will introduce the model components in this section, and detailed configurations are posted in Appendix \ref{app:setting_details}.

\vspace{-3pt}
\paragraph{Encoder.}
The encoder begins with a one-dimensional convolutional projection, followed by 6 strided causal convolutional downsampling blocks. Across these blocks, the channel width doubles from 12 to 768, with kernel sizes \(4,4,4,8,12,20\) and downsampling rates \(2,2,2,4,6,10\). This gives a total hop size of 1920, corresponding to a 25 Hz latent sequence for 48 kHz audio. Each downsampling block is followed by a residual stack of dilated causal convolutions. In our configuration, each stack contains 6 residual layers, which enlarge the receptive field while preserving causal processing. The final encoder projection maps the hidden sequence to a 128-dimensional acoustic representation. To improve reconstruction quality under a bounded-latency constraint, the encoder is causal except for a final 2-frame lookahead convolution.

\vspace{-3pt}
\paragraph{Temporal variational bottleneck.}
On top of the convolutional encoder, we add bottleneck layers, consisting of a 4-layer LSTM block with project-in and -out linear layers. A \(1\times1\) convolution then predicts the mean and log-scale of a diagonal Gaussian posterior, from which the latent sequence is sampled via the reparameterization trick. To increase the expressiveness of the latent distribution, we further apply a normalizing flow when computing the KL regularization against the standard normal prior~\cite{kalle, flowvae}. The sampled latent sequence is projected back to the model dimension and processed by a mirrored structure of encoder-side bottleneck before decoding. 

\vspace{-3pt}
\paragraph{Decoder.}
The decoder reconstructs the 48k~Hz waveform from the 25 Hz latent sequence using a BigVGAN-style generator. Its upsampling module mirrors the encoder downsampling structure. Differently, following BigVGAN, each upsampling stage is refined by AMPBlocks with SnakeBeta activation. Similar to the encoder, the decoder introduces a 2-frame lookahead in its first convolutional net and is otherwise causal. The final projection maps the hidden features to a single-channel waveform.

\vspace{-3pt}
\paragraph{Supervision network.}
The role of this component is detailed in Section~\ref{sec:latent_distill_supervise}. The supervision network follows an encoder--decoder design, consisting of a 0.6B Transformer encoder and a pretrained Qwen2.5-0.5B~\cite{qwen2025qwen25technicalreport} decoder. The encoder processes latent samples, which are concatenated with task-label embeddings and passed to the language-model decoder. This decoder is used only for Stage-III supervision and discarded afterwards; downstream generation always uses the frozen waveform decoder.


\vspace{-4.5pt}
\subsection{Stage I\&II: Progressive Training of High-fidelity Variational Latent Space}
\label{sec:training_method}

Empirically, imposing a strong KL constraint in VAE training can promote a more structured latent distribution, but it may also force the representation to discard acoustic details before the decoder has learned a high-fidelity reconstruction manifold. To mitigate this fidelity loss, we progressively shape the \ourmethod{} latent space instead of learning it in a single stage. The overview is shown as Figure \ref{fig:overview}. Stage I trains a deterministic autoencoder to establish a high-fidelity acoustic autoencoding space. Stage II freezes the pretrained encoder and decoder, and converts this autoencoding space into a stochastic latent space by training only a temporal variational bottleneck with weak KL regularization. This staged procedure keeps the latent trajectory close to a reliable decoding region, providing a stable foundation for downstream-aware Stage III training. We further analyze this process as implicit fidelity transfer.

\vspace{-3pt}
\paragraph{Stage I: reconstruction-oriented autoencoder pretraining.}
Given an input waveform \(\mathbf{x}\), the encoder \(E_{\phi}\) maps it to a low-rate acoustic representation,
\(
\mathbf{z}_{\mathrm{AE}} = E_{\phi}(\mathbf{x}),
\)
from which the decoder \(G_{\psi}\) reconstructs the waveform as
\(
\hat{\mathbf{x}}_{\mathrm{AE}} = G_{\psi}(\mathbf{z}_{\mathrm{AE}}).
\)
This stage is trained with a reconstruction-oriented generator objective:
\vspace{-4pt}
\begin{equation}
\begin{aligned}
\mathcal{L}_{\mathrm{I}}
=
\mathbb{E}_{\mathbf{x}}
\left[
\ell_{\mathrm{gen}}
\bigl(
\mathbf{x},
G_{\psi}(E_{\phi}(\mathbf{x}))
\bigr)
\right],
\end{aligned}
\end{equation}
\vspace{-3pt}
where \(\ell_{\mathrm{gen}}\) denotes the generator-side waveform generation loss, combining multi-scale spectral reconstruction, adversarial supervision, and discriminator feature matching:
\vspace{-4pt}
\begin{equation}
\ell_{\mathrm{gen}}
=
\lambda_{\mathrm{spec}}\mathcal{L}_{\mathrm{spec}}
+
\lambda_{\mathrm{adv}}\mathcal{L}_{\mathrm{adv}}^{G}
+
\lambda_{\mathrm{fm}}\mathcal{L}_{\mathrm{fm}} .
\end{equation}
\vspace{-3pt}
Here, \(\mathcal{L}_{\mathrm{spec}}\) is the multi-scale mel-spectral reconstruction loss, \(\mathcal{L}_{\mathrm{adv}}^{G}\) is the generator-side adversarial loss, and \(\mathcal{L}_{\mathrm{fm}}\) is the feature matching loss computed from discriminator intermediate activations. The discriminator objective is optimized in parallel and omitted for notational clarity. This stage establishes a high-fidelity reconstruction manifold before introducing variational regularization.

\vspace{-3pt}
\paragraph{Stage II: autoencoding-to-variational latent transfer.}
Starting from the pretrained autoencoder, we freeze the encoder \(E_{\phi}\) and decoder \(G_{\psi}\), and train only the temporal variational bottleneck. Given the deterministic acoustic representation
\(
\mathbf{z}_{\mathrm{AE}} = E_{\phi}(\mathbf{x}),
\)
the bottleneck defines a posterior
\(
q_{\eta}(\mathbf{z}_{\mathrm{VAE}}|\mathbf{z}_{\mathrm{AE}})
\)
over stochastic latents, which are sampled with the reparameterization trick and decoded by the frozen decoder. We optimize a reconstruction-dominated VAE objective:
\vspace{-4pt}
\begin{equation}
\begin{aligned}
\mathcal{L}_{\mathrm{II}}
&=
\mathbb{E}_{\mathbf{x}}
\Bigg[
\mathbb{E}_{\mathbf{z}_{\mathrm{VAE}}\sim q_{\eta}(\cdot|\mathbf{z}_{\mathrm{AE}})}
\left[
\ell_{\mathrm{gen}}
\bigl(
\mathbf{x},G_{\psi}(\mathbf{z}_{\mathrm{VAE}})
\bigr)
\right]
\\
&\quad+
\beta_{\mathrm{low}}
D_{\mathrm{KL}}
\left(
q_{\eta}(\mathbf{z}_{\mathrm{VAE}}|\mathbf{z}_{\mathrm{AE}})
\|p(\mathbf{z})
\right)
\Bigg],
\end{aligned}
\end{equation}
\vspace{-3pt}
where \(p(\mathbf{z})=\mathcal{N}(\mathbf{0},\mathbf{I})\). The small KL weight encourages distributional regularity without forcing the bottleneck to discard reconstruction-critical acoustic details. Since \(E_{\phi}\) and \(G_{\psi}\) remain fixed, Stage II transfers the deterministic autoencoding space into a variational latent space while keeping sampled latents close to the decoder's high-fidelity reconstruction region.

\vspace{-3pt}
\paragraph{Implicit fidelity transfer.}
The progressive Stage-I/II design provides an implicit fidelity-transfer effect. As formalized in Appendix~\ref{app:proof_fidelity_transfer}, the frozen pretrained decoder and the reconstruction-dominated objective constrain Stage-II variational samples to stay near the high-fidelity autoencoding manifold, so their expected waveform distortion is controlled by the Stage-I autoencoder distortion and the AE-to-VAE latent shift. This motivates training only the temporal variational bottleneck with a small KL weight while keeping the pretrained decoder frozen. 

\vspace{-4.5pt}
\subsection{Stage III: Downstream-aware Enrichment of the Tokenization Space}
\label{sec:latent_distill_supervise}

After Stages I--II, the latent space has acquired high-fidelity reconstruction ability and initial variational regularity. However, reconstruction alone does not guarantee that the latent sequence preserves information required by downstream understanding tasks. In Stage III, we further enrich the VAE latent space with pretrained speech representations and task-conditioned supervision, making the tokenization space both waveform-decodable and informative for downstream speech-language modeling. We denote the full VAE posterior by
\(
q_{\theta}(\mathbf{z}|\mathbf{x})
=
q_{\eta'}(\mathbf{z}|E_{\phi'}(\mathbf{x}))
\),
which inherits the same bottleneck architecture as the Stage-II posterior
\(
q_{\eta}(\mathbf{z}_{\mathrm{VAE}}|E_{\phi}(\mathbf{x}))
\)
and is initialized from it. The new notation emphasizes that the encoder and bottleneck are jointly optimized during Stage III.

\vspace{-3pt}
\paragraph{Multi-granularity representation distillation.}
We introduce multi-granularity representation distillation to enrich the VAE latent space beyond waveform reconstruction. Given \(\mathbf{z}\sim q_{\theta}(\mathbf{z}|\mathbf{x})\), we align the latent sequence with frozen teacher representations from pretrained speech models at both frame and utterance levels. For frame-level distillation, following~\cite{niu2025semanticvae}, we use WavLM~\cite{wavlm} as a contextual teacher and apply a prediction head to map the latent sequence to its 23rd-layer hidden representations, with temporal interpolation used when frame rates differ. For utterance-level distillation, we aggregate the latent sequence into an utterance-level representation and align it with an x-vector speaker embedding~\cite{ecapa-tdnn}. The unified distillation objective is
\vspace{-2pt}
\begin{equation}
\resizebox{.98\columnwidth}{!}{%
$
\begin{aligned}
\mathcal{L}_{\mathrm{distill}}
=
\sum_{r\in\mathcal{R}}
\lambda_{r}
\left[
1-
\cos
\left(
H_{r}(A_{r}(\mathbf{z})),
\operatorname{sg}(F_{r}(\mathbf{x}))
\right)
\right],
\end{aligned}
$%
}
\label{eq:distill}
\end{equation}
\vspace{-2pt}
where \(\mathcal{R}\) denotes the set of teacher representations, \(F_{r}\) is a frozen teacher, \(A_{r}\) performs temporal alignment for frame-level teachers or pooling for utterance-level teachers, \(H_{r}\) maps the adapted latent representation to the teacher space, and \(\operatorname{sg}(\cdot)\) denotes stop-gradient. For frame-level teachers, the cosine term is computed after temporal alignment and averaged over time.

\vspace{-3pt}
\paragraph{Multi-task language-modeling supervision.}
We further expose the latent representation to downstream supervision through the task-conditioned supervision network described in Section~\ref{sec:architecture}. Given a task type \(\tau\in\mathcal{T}\) and its target output \(\mathbf{y}^{\tau}\), we optimize a unified language-modeling objective:
\vspace{-2pt}
\begin{equation}
\begin{aligned}
\mathcal{L}_{\mathrm{sup}}
=
-
\mathbb{E}_{(\mathbf{x},\tau,\mathbf{y}^{\tau})}
\mathbb{E}_{\mathbf{z}\sim q_{\theta}(\cdot|\mathbf{x})}
\left[
\log p_{\omega}(\mathbf{y}^{\tau}\mid\mathbf{z},\tau)
\right].
\end{aligned}
\label{eq:sup}
\end{equation}
\vspace{-2pt}
This formulation converts heterogeneous downstream annotations into a shared task-conditioned prediction interface, covering tasks including speech recognition, emotion recognition, audio captioning, and sound event detection. As a result, the latent space is encouraged to retain information that may be unnecessary for waveform reconstruction but is critical for speech and audio understanding.

Combining waveform reconstruction, variational regularization, representation distillation, and downstream supervision, the Stage-III objective is
\vspace{-2pt}
\begin{equation}
\mathcal{L}_{\mathrm{III}}
=
\mathcal{L}_{\mathrm{gen}}
+
\beta_{\mathrm{high}}\mathcal{L}_{\mathrm{KL}}
+
\mathcal{L}_{\mathrm{distill}}
+
\lambda_{\mathrm{sup}}\mathcal{L}_{\mathrm{sup}} .
\end{equation}
\vspace{-2pt}
Here, \(\mathcal{L}_{\mathrm{gen}}\) denotes the expected generator-side waveform generation loss, \(\mathcal{L}_{\mathrm{KL}}\) regularizes the VAE posterior toward the standard normal prior, and \(\beta_{\mathrm{high}}\) is much larger than the weak KL weight used in Stage II.

\vspace{-3pt}
\paragraph{Variational interpretation.}
Stage III can be interpreted as optimizing a downstream-aware variational surrogate. Let \(\mathbf{u}_{r}=F_{r}(\mathbf{x})\) denote a frozen teacher representation. We view the latent variable \(\mathbf{z}\) as jointly explaining the waveform, teacher representations, and task target:
\vspace{-2pt}
\begin{equation}
\resizebox{.98\columnwidth}{!}{%
$
p(\mathbf{x},\{\mathbf{u}_{r}\}_{r\in\mathcal{R}},\mathbf{y}^{\tau}\mid\tau)
=
\int
p(\mathbf{z})p_{\psi}(\mathbf{x}\mid\mathbf{z})
p_{\omega}(\mathbf{y}^{\tau}\mid\mathbf{z},\tau)
\prod_{r\in\mathcal{R}}p_{r}(\mathbf{u}_{r}\mid\mathbf{z})
\,d\mathbf{z}.
$%
}
\end{equation}
\vspace{-2pt}
With the variational posterior \(q_{\theta}(\mathbf{z}\mid\mathbf{x})\), this gives the weighted ELBO-style objective
\vspace{-2pt}
\begin{equation}
\resizebox{.98\columnwidth}{!}{%
$
\begin{aligned}
\mathcal{J}_{\mathrm{III}}
&=
\mathbb{E}_{\mathbf{z}\sim q_{\theta}(\cdot|\mathbf{x})}
\Big[
\log p_{\psi}(\mathbf{x}\mid\mathbf{z})
+
\lambda_{\mathrm{sup}}
\log p_{\omega}(\mathbf{y}^{\tau}\mid\mathbf{z},\tau)
\\
&\quad+
\sum_{r\in\mathcal{R}}
\lambda_{r}
\log p_{r}(\mathbf{u}_{r}\mid\mathbf{z})
\Big]
-
\beta_{\mathrm{high}}
D_{\mathrm{KL}}
\left(
q_{\theta}(\mathbf{z}\mid\mathbf{x})
\|p(\mathbf{z})
\right).
\end{aligned}
$%
}
\end{equation}
\vspace{-2pt}
Minimizing \(\mathcal{L}_{\mathrm{III}}\) can therefore be viewed as maximizing this surrogate with practical waveform, distillation, supervision, and KL terms.

\begin{table*}[!t]
\centering
\small
\setlength{\tabcolsep}{4pt}
\renewcommand{\arraystretch}{1.12}
\begin{tabular}{@{}lcccccccc@{}}
\toprule
\textbf{Model} & \textbf{CR} & \textbf{TPS} & \textbf{NB/WB PESQ} $\uparrow$ & \textbf{STOI} $\uparrow$ & \textbf{WER(\%)} $\downarrow$ & \textbf{SPKSIM} $\uparrow$ & \textbf{EMOSIM} $\uparrow$ & \textbf{UTMOS} $\uparrow$ \\
\midrule
\midrule
Ground Truth & 1.00$\times$ & -- & -- & -- & 3.91 & 1.000 & 1.000 & 3.75 \\
\midrule
Mel Spectrogram & 2.00$\times$ & 86 & 4.15/4.05 & \textbf{0.988} & \textbf{3.96} & 0.957 & 0.988 & \underline{3.75}  \\
SemanticVAE & 2.73$\times$ & 40 & 3.99/3.80 & 0.969 & \underline{4.15} & 0.963 & \underline{0.993} & \textbf{3.76} \\
MingTok-Audio & 2.19$\times$ & 50 & \textbf{4.23/4.12} & 0.981 & 4.27 & 0.950 & 0.992 & \underline{3.75} \\
\midrule
Vanilla VAE & \multirow{2}{*}{\textbf{7.5$\times$}} & \multirow{2}{*}{25} & 3.18/2.65 & 0.925 & 5.41 & 0.859 & 0.988 & \underline{3.75} \\
\ourmethod{} & & & \underline{4.10/4.01} & \underline{0.974} & 4.22 & \textbf{0.968} & \textbf{0.995} & \underline{3.75} \\
\bottomrule
\end{tabular}
\caption{Reconstruction evaluation results on LibriSpeech test-other.}
\label{tab:reconstruction-evaluation}
\vspace{-9pt}
\end{table*}

\vspace{-4.5pt}
\subsection{Downstream Unified Spoken Language Modeling}
\label{sec:downstream_model}

To evaluate whether the learned speech representation can serve as a unified modeling space, we build a downstream spoken language model that supports both speech understanding and speech generation with a shared backbone. Inspired by ~\cite{jia2025ditar}, the model follows an AR+DiT design: an autoregressive language model processes mixed text--audio embedding sequences, while a DiT-based flow-matching module predicts continuous latent patches for speech generation. The architecture overview is on right side of Figure \ref{fig:overview}.

\vspace{-3pt}
\paragraph{Speech understanding objective.}
Let \(\mathbf{z}_{\mathrm{audio}}\) denote the audio latent patches and \(\mathbf{e}_{\mathrm{audio}}\) denote their corresponding language-model embeddings. Given textual context \(\mathbf{c}\) and target text \(\mathbf{y}_{\mathrm{text}}\), we optimize an autoregressive cross-entropy objective:
\vspace{-2pt}
\begin{equation}
\mathcal{L}_{\mathrm{understand}}
=
-
\sum_{j}
\log p_{\theta}
\left(
y_{j}
\mid
\mathbf{y}_{<j},
\mathbf{e}_{\mathrm{audio}},
\mathbf{c}
\right).
\end{equation}
\vspace{-2pt}

\vspace{-3pt}
\paragraph{Speech generation objective.}
For speech generation, the autoregressive language model summarizes the available text and audio history into causal hidden states, and the DiT flow-matching module predicts each future latent patch conditioned on this previous hiddens and historical latents with an autoregressive pattern following \cite{liu2024ardit}. The conditional generation process is factorized as
\vspace{-2pt}
\begin{equation}
p_{\theta}
\left(
\mathbf{z}_{1:K}
\mid
\mathbf{c}
\right)
=
\prod_{k=1}^{K}
p_{\theta}
\left(
\mathbf{z}_{k}
\mid
\mathbf{h}_{\le k}, \mathbf{z}_{<k}
\right),
\end{equation}
\vspace{-2pt}
where \(\mathbf{h}_{\le k}\) is the causal language-model hidden states sequence for \(k\) patches prediction, and \(\mathbf{z}_{<k}\) denotes previously generated audio latent patches. Each conditional patch distribution is learned with a flow-matching objective:
\vspace{-2pt}
\begin{equation}
\mathcal{L}_{\mathrm{FM}}
=
\mathbb{E}_{k,t}
\left[
\left\|
v_{\theta}
\left(
\mathbf{z}_{k,t}, t
\mid
\mathbf{h}_{\le k}, \mathbf{z}_{<k}
\right)
-
\mathbf{u}_{k,t}
\right\|_{2}^{2}
\right],
\end{equation}
\vspace{-2pt}
where \(\mathbf{z}_{k,t}\) is the interpolated noisy state of the \(k\)-th latent patch at timestamp \(t\), and \(\mathbf{u}_{k,t}\) is the corresponding target velocity. We further supervise audio termination with a binary cross-entropy EOS loss:
\vspace{-2pt}
\begin{equation}
\mathcal{L}_{\mathrm{generate}}
=
\mathcal{L}_{\mathrm{FM}}
+
\lambda_{\mathrm{eos}}
\mathcal{L}_{\mathrm{eos}}.
\end{equation}
\vspace{-2pt}
The generated latent patches are assembled into a latent sequence and decoded into waveform audio by the frozen \ourmethod{} decoder.

%% file: experiment/exp.tex
\section{Experiments}
\label{sec:exp}
\vspace{-5pt}
\subsection{Experimental settings and Baselines}

\vspace{-3pt}
\paragraph{Training datasets.}
We train \ourmethod{} on a mixture of speech, environmental sound, and music data.
The speech data include AISHELL-3~\cite{shi21c_interspeech},
HiFi-TTS~\cite{bakhturina21_interspeech},
VCTK~\cite{yamagishi2019vctk},
HiFiTTS2~\cite{langman25_interspeech},
and large-scale internal English and Chinese TTS corpora, totaling approximately 500K hours.
To improve robustness beyond clean read speech, we further include emotional speech data,
AudioSet~\cite{7952261},
VGGSound~\cite{9053174},
VocalSound~\cite{Gong_2022},
FSD50K~\cite{9645159},
MusicCaps~\cite{agostinelli2023musiclmgeneratingmusictext},
and WavCaps~\cite{10572302}.

\vspace{-3pt}
\paragraph{Training settings.}
All audio is resampled to 48~kHz. Following the BigVGAN v2 configuration~\cite{lee2023bigvgan}, the generator uses a multi-period discriminator and a multi-scale sub-band CQT discriminator. Training proceeds in three stages: 500K steps for autoencoder pretraining, 50K steps for the variational bottleneck with \(\beta_{\mathrm{low}}=0.1\), and 200K steps for Stage-III full-model training with \(\beta_{\mathrm{high}}=7\). Detailed optimization settings are provided in Appendix~\ref{app:setting_details}.

\vspace{-3pt}
\paragraph{Main baselines.}
We compare \ourmethod{} with two representative continuous audio representations. Semantic-VAE~\cite{niu2025semanticvae} distills pretrained SSL representations into VAE latents and has shown strong performance for DiT-based speech synthesis over mel-spectrogram inputs. MingTok-Audio is a continuous speech tokenizer designed for AR+DiT-based unified speech understanding and generation. For MingTok-Audio, we use its unified feature as the input representation and its acoustic latent as the generation target, while keeping the semantic module fixed following its reported ablation protocol~\cite{yan2025minguniaudiospeechllmjoint}.

\vspace{-5pt}
\subsection{Reconstruction Evaluation}

We evaluate reconstruction quality on LibriSpeech~\cite{librispeech} test-other in terms of signal fidelity, linguistic preservation, and paralinguistic consistency. We report narrow-band and wide-band PESQ~\cite{pesq}, STOI~\cite{stoi}, and UTMOS~\cite{utmos} for perceptual quality and intelligibility; WER on resynthesized speech for linguistic preservation; and speaker similarity(SPKSIM)~\cite{ecapa-tdnn} and emotion similarity(EMOSIM)~\cite{emotion2vec} for paralinguistic consistency. Ground-truth waveforms are used as references for signal-level metrics. We compare \ourmethod{} with BigVGAN v2 mel-spectrogram vocoding\footnote{\href{https://huggingface.co/nvidia/bigvgan_v2_44khz_128band_512x}{BigVGAN v2 checkpoint}}, directly trained VAE and main baselines.

We also report tokens per second (TPS) and compression ratio (CR). CR is computed as the ratio between the raw waveform nominal bitrate and the latent representation bitrate, indicating the real information compression rate:
\begin{equation}
\mathrm{CR}
=
(f_s \left\lceil \log_2 f_s \right\rceil)
/(f_z d_z b_{\mathrm{float}}),
\end{equation}
where \(f_s\) is the waveform sampling rate, \(\lceil \log_2 f_s \rceil\) is the norminal number of bits used for each waveform sample, \(f_z\) is the latent frame rate, \(d_z\) is the latent dimension, and \(b_{\mathrm{float}}=32\) is the number of bits per floating-point latent value.

As shown in Table~\ref{tab:reconstruction-evaluation}, \ourmethod{} achieves competitive reconstruction quality among continuous speech representations while using the most compact latent sequence, with a compression ratio of 7.5\(\times\) and 25 TPS. Although mel-spectrogram vocoding and MingTok-Audio obtain slightly higher scores on some signal-level metrics, \ourmethod{} preserves linguistic and paralinguistic information well, achieving strong WER, the best SPKSIM, and the best EMOSIM. Compared with the vanilla VAE using the same architecture and compression rate, \ourmethod{} substantially improves PESQ, STOI, WER, and SPKSIM, validating the effectiveness of the progressive training strategy. Subjective reconstruction scores are reported in Appendix~\ref{app:setting_details}.


\vspace{-5pt}
\subsection{Evaluation on Speech Synthesis}

Speech synthesis directly tests whether a representation is learnable as a generation target. We use the generation branch of the AR+DiT model in Section~\ref{sec:downstream_model}. A base TTS model is trained on 95K hours of filtered Emilia~\cite{emilia} data for 200k steps, and then further tuned 50k steps for controllable TTS with EmoVoice-DB~\cite{emovoice}, FCaps~\cite{clsp}, and PSCBase~\cite{PSC-data}. Details are given in Appendix~\ref{app:setting_details}.

\vspace{-3pt}
\paragraph{Zero-shot capability and synthesis diversity.}
We evaluate the base model in a zero-shot setting, synthesizing unseen-speaker speech from prompt speech and text. On Seed-TTS-Eval~\cite{seedtts}, we report WER and speaker similarity for intelligibility and speaker preservation, respectively. We further evaluate the emotion and paralinguistic subsets of Emergent-TTS~\cite{manku2026emergentttseval}, reporting WER and win rate against GPT-4o-mini-TTS~\cite{openai2024gpt4ocard}. Table~\ref{tab:zero_shot_tts} shows that \ourmethod{} achieves competitive zero-shot TTS performance and obtains the highest win rates on both expressive dimensions, suggesting that its latent space is highly learnable and preserves expressive paralinguistic information.



\begin{table*}[t]
\centering
\small
\setlength{\tabcolsep}{2pt}
\renewcommand{\arraystretch}{1.12}
\resizebox{\textwidth}{!}{%
\begin{tabular}{@{}ccccccccccc@{}}
\toprule
\multirow{3}{*}{\textbf{Model}} & \multicolumn{6}{c}{\textbf{Seed-TTS-Eval}} & \multicolumn{4}{c}{\textbf{Emergent-TTS}} \\
\cmidrule(lr){2-7}\cmidrule(l){8-11}
 & \multicolumn{2}{c}{\textbf{Seed-TTS-en}} & \multicolumn{2}{c}{\textbf{Seed-TTS-zh}} & \multicolumn{2}{c}{\textbf{Seed-TTS-hard}} & \multicolumn{2}{c}{\textbf{Emotion}} & \multicolumn{2}{c}{\textbf{Paralinguistic}} \\
\cmidrule(lr){2-3}\cmidrule(lr){4-5}\cmidrule(lr){6-7}\cmidrule(lr){8-9}\cmidrule(l){10-11}
 & \textbf{WER(\%)} $\downarrow$ & \textbf{SIM} $\uparrow$ & \textbf{WER(\%)} $\downarrow$ & \textbf{SIM} $\uparrow$ & \textbf{WER(\%)} $\downarrow$ & \textbf{SIM} $\uparrow$ & \textbf{WER(\%)} $\downarrow$ & \textbf{Win-Rate} $\uparrow$ & \textbf{WER(\%)} $\downarrow$ & \textbf{Win-Rate} $\uparrow$ \\
\midrule
Semantic-VAE & \underline{1.42} & \textbf{0.63} & \textbf{0.91} & \underline{0.70} & \textbf{7.53} & \textbf{0.67} & \textbf{0.63} & \underline{14.3} & \textbf{34.45} & \underline{44.2} \\
MingTok-Audio & 1.84 & 0.61 & 1.03 & \textbf{0.71} & 14.75 & \underline{0.66} & 8.81 & 8.4 & 35.66 & 39.8 \\
\ourmethod{} & \textbf{1.33} & \underline{0.62} & \underline{0.98} & \underline{0.70} & \underline{7.59} & \underline{0.66} & \underline{1.34} & \textbf{25.5} & \underline{34.47} & \textbf{53.6} \\
\bottomrule
\end{tabular}%
}
\caption{Zero-shot TTS evaluation on Seed-TTS-Eval and Emergent-TTS.``SIM'' refers to the speaker similarity between synthesized and prompt speech.}
\label{tab:zero_shot_tts}
\vspace{-8pt}
\end{table*}

\begin{figure}[t]
\centering
\includegraphics[width=\columnwidth]{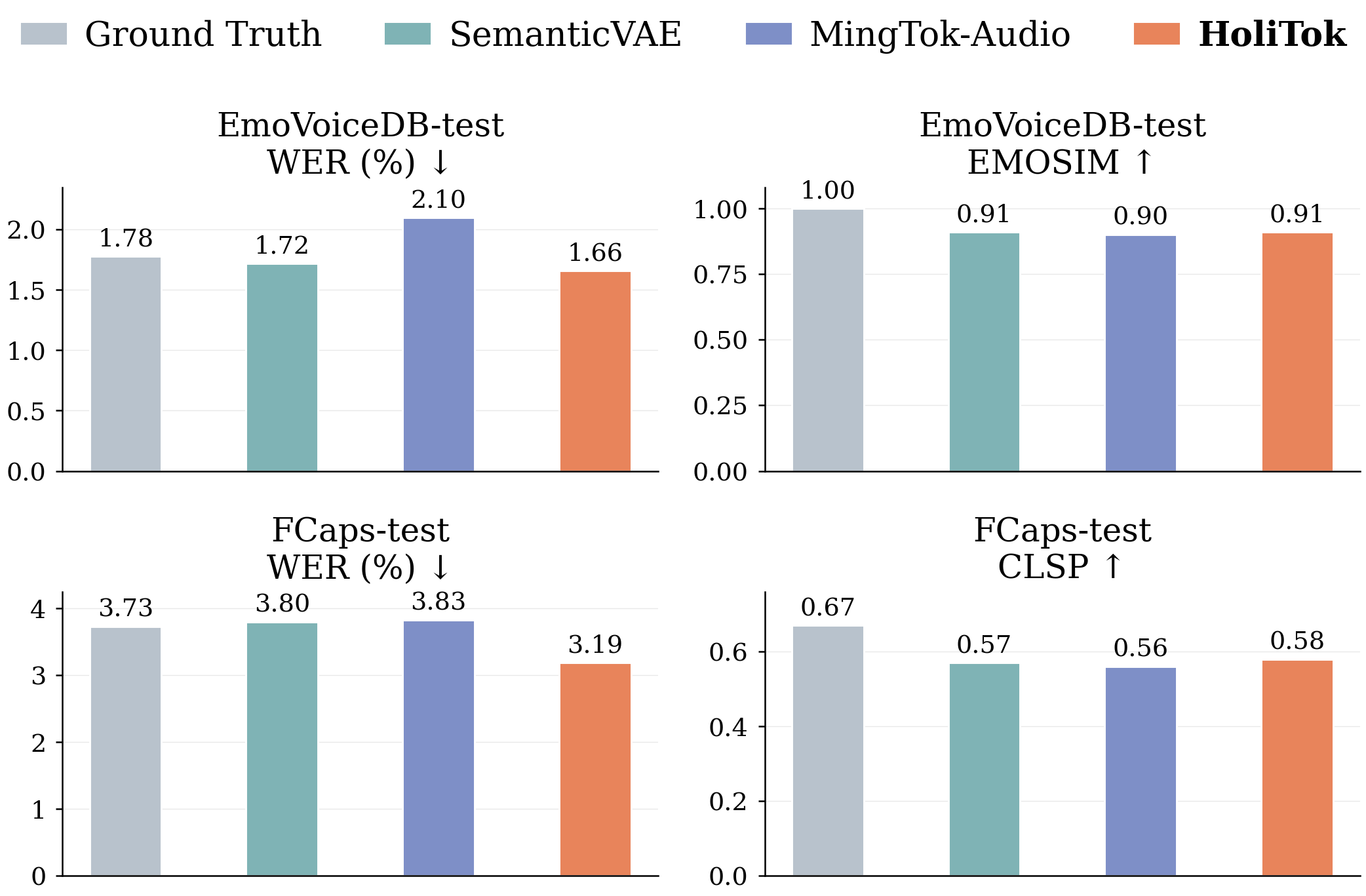}
\caption{Controllable TTS evaluation on EmoVoiceDB-test and FCaps-test.}
\label{fig:controllable_tts}
\vspace{-10pt}
\end{figure}

\vspace{-3pt}
\paragraph{Controllable TTS.}
We evaluate the fine-tuned TTS model on controllable synthesis, where speech is generated from explicit emotional or paralinguistic descriptions. On EmoVoiceDB-test, we report WER and EMOSIM for content consistency and emotion control. On FCaps-test, we report WER and CLSP~\cite{clsp} score to measure alignment with fine-grained speaking-style descriptions. As shown in Figure~\ref{fig:controllable_tts}, \ourmethod{} achieves the best WER on both datasets, matches the best EMOSIM, and obtains the highest CLSP score, indicating stronger controllability without sacrificing intelligibility.

\begin{table*}[!t]
\centering
\small
\setlength{\tabcolsep}{4pt}
\renewcommand{\arraystretch}{1.12}
\resizebox{\textwidth}{!}{%
\begin{tabular}{@{}cccccccccc@{}}
\toprule
\multirow{3}{*}{\textbf{Model}} & \multicolumn{6}{c}{\textbf{TTS}} & \multicolumn{3}{c}{\textbf{ASR}} \\
\cmidrule(lr){2-7}\cmidrule(l){8-10}
 & \multicolumn{2}{c}{\textbf{Seed-TTS-en}} & \multicolumn{2}{c}{\textbf{Seed-TTS-zh}} & \multicolumn{2}{c}{\textbf{Seed-TTS-hard}} & \textbf{test-clean} & \textbf{test-other} & \textbf{AISHELL-1} \\
\cmidrule(lr){2-3}\cmidrule(lr){4-5}\cmidrule(lr){6-7}\cmidrule(l){8-10}
 & \textbf{WER(\%)} $\downarrow$ & \textbf{SIM} $\uparrow$ & \textbf{WER(\%)} $\downarrow$ & \textbf{SIM} $\uparrow$ & \textbf{WER(\%)} $\downarrow$ & \textbf{SIM} $\uparrow$ & \textbf{WER(\%)} $\downarrow$ & \textbf{WER(\%)} $\downarrow$ & \textbf{WER(\%)} $\downarrow$ \\
\midrule
Semantic-VAE & 102.32 & 0.47 & 99.30 & 0.61 & 97.31 & \underline{0.61} & 9.69 & 21.32 & 15.81 \\
MingTok-Audio & 51.06 & 0.42 & 18.17 & 0.61 & 50.35 & 0.55 & \textbf{4.62} & \textbf{9.06} & \textbf{5.01} \\
\ourmethod{}-Base & \underline{27.85} & \underline{0.52} & \underline{4.40} & \underline{0.66} & \underline{30.44} & \underline{0.61} & 6.45 & 16.51 & 14.92 \\
\ourmethod{}-Unite & \textbf{7.20} & \textbf{0.55} & \textbf{1.78} & \textbf{0.67} & \textbf{16.79} & \textbf{0.62} & \underline{5.48} & \underline{12.65} & \underline{5.93} \\
\bottomrule
\end{tabular}%
}
\caption{Unified spoken language modeling evaluation on TTS and ASR tasks.}
\label{tab:unified_slm}
\vspace{-10pt}
\end{table*}

\vspace{-5pt}
\subsection{Evaluation on Unified Understanding and Generation}

\vspace{-3pt}
\paragraph{Settings.}
We use the AR+DiT architecture in Section~\ref{sec:downstream_model} for unified spoken language modeling, instantiating understanding as ASR and generation as TTS. \ourmethod{}-Base uses the learned VAE latents as the audio representation; its non-causal supervision encoder is used only during representation training. \ourmethod{}-Unite uses the causal supervision encoder trained in Stage III as a built-in semantic encoder, replacing the downstream patch encoder and providing pre-modeled speech features, similar in spirit to MingTok-Audio. We train the unified model with Emilia for TTS and AISHELL-1/2~\cite{aishell1, aishell2}, GigaSpeech~\cite{gigaspeech}, MLS~\cite{mls}, Common Voice 20.0~\cite{commonvoice}, FLEURS~\cite{fleurs}, and LibriSpeech~\cite{librispeech} for ASR, using a sampler that keeps the TTS-to-ASR ratio near 5:1. We evaluate TTS on Seed-TTS-Eval and ASR on LibriSpeech test-clean/test-other and AISHELL-1 test.

\vspace{-3pt}
\paragraph{Evaluation results analysis.}
Table~\ref{tab:unified_slm} shows that unified ASR--TTS training is substantially more demanding than task-specific modeling in Table~\ref{tab:zero_shot_tts}. Under the same AR+DiT architecture, existing continuous representations degrade sharply on TTS, indicating that reconstruction quality or isolated downstream performance does not necessarily translate to usability in a shared generation-understanding model. Within this unified setting, \ourmethod{} shows a better balance between generation and understanding. \ourmethod{}-Base already outperforms the baselines on all TTS intelligibility and achieves comparable ASR results. This suggests that the proposed VAE latent space remains more learnable as a continuous generation target while preserving sufficient acoustic information. With the causal semantic encoder, \ourmethod{}-Unite further reduces the average TTS WER from 20.90\% to 8.59\% and improves the average ASR WER from 12.63\% to 8.02\% over \ourmethod{}-Base. These gains indicate that the Stage-III causal encoder provides useful pre-learning of \ourmethod{} representations, rather than merely improving an isolated understanding branch.
The comparison also reveals different failure modes of existing representations. Semantic-VAE obtains usable ASR performance but fails on TTS, suggesting that directly shaping the latent space toward semantic representations can weaken its generative learnability. MingTok-Audio achieves the best ASR WER, but its TTS performance remains much weaker than \ourmethod{}-Unite, indicating an imbalance toward understanding. Overall, \ourmethod{} better satisfies the joint requirements of unified spoken language modeling: acoustic preservation for generation, semantic accessibility for understanding, and latent learnability under a shared AR+DiT backbone.

\vspace{-3pt}
\paragraph{Ablation study.}
We provide complete ablation results in Appendix~\ref{app:unified_ablation}, showing that the proposed training components and downstream modeling choices are complementary. Only high variational regularization alone is insufficient for a generation-friendly unified representation. Using representation distillation without supervision also severely degrades synthesis, consistent with the Semantic-VAE results, while supervision alone preserves much stronger TTS performance. On the downstream side, DiT initialization with TTS-only training consistently improves generation quality, and \ourmethod{}-Unite performs best when the causal semantic encoder remains trainable rather than frozen.

%% file: appendix.tex
\appendix
\begin{table*}[t]
\centering
\small
\setlength{\tabcolsep}{4pt}
\renewcommand{\arraystretch}{1.08}
\resizebox{\textwidth}{!}{%
\begin{tabular}{@{}lcccc@{}}
\toprule
\textbf{Configuration} &
\textbf{\ourmethod{}-Base} &
\textbf{\ourmethod{}-Unite} &
\textbf{SemanticVAE} &
\textbf{MingTok-Audio} \\
\midrule
Audio representation & VAE latent & semantic feature & VAE latent & semantic feature \\
Input dimension & 128 & 1536 & 64 & 1280 \\
Temporal aggregation & patch size 4 & mean pooling size 4 & patch size 4 & mean pooling size 8 \\
Audio-to-LLM mapping & 8-layer PatchEncoder & linear projection & 8-layer PatchEncoder & linear projection \\
Audio-to-LLM params & 102M & 1M & 102M & 1M \\
LLM backbone & Qwen2.5-0.5B & Qwen2.5-0.5B & Qwen2.5-0.5B & Qwen2.5-0.5B \\
LLM hidden size & 896 & 896 & 896 & 896 \\
LLM params & 494M & 494M & 494M & 494M \\
DiT layers & 18 & 18 & 18 & 18 \\
DiT hidden size & 1024 & 1024 & 1024 & 1024 \\
DiT attention heads & 16 & 16 & 16 & 16 \\
DiT FFN hidden size & 4096 & 4096 & 4096 & 4096 \\
DiT input/output dim. & 1024 / 128 & 1024 / 128 & 1024 / 64 & 1024 / 64 \\
DiT params & 345M & 345M & 345M & 345M \\
Bridge / EOS params & 2M & 2M & 2M & 2M \\
\midrule
Total downstream params & 942M & 842(+680)M & 942M & 841M \\
\bottomrule
\end{tabular}
}
\caption{Downstream AR+DiT architecture configuration and parameter counts. Tokenizer-side modules in Table~\ref{tab:tokenizer_config} are not included.}
\label{tab:downstream_arch_config}
\end{table*}

\begin{table}[t]
\centering
\small
\setlength{\tabcolsep}{4pt}
\renewcommand{\arraystretch}{1.08}
\resizebox{\columnwidth}{!}{%
\begin{tabular}{@{}ll@{}}
\toprule
\textbf{Symbol} & \textbf{Template} \\
\midrule
\(\mathcal{P}_{\mathrm{tts}}\) &
\([\mathrm{text}]\,\mathbf{t}\,[\mathrm{text2speech}]\,\langle\mathrm{loss}\rangle\mathbf{a}\langle\mathrm{eos}\rangle\langle/\mathrm{loss}\rangle\) \\
\(\mathcal{P}_{\mathrm{asr}}\) &
\([\mathrm{speech}]\,\mathbf{a}\,[\mathrm{speech2text}]\,\langle\mathrm{loss}\rangle\mathbf{t}\langle\mathrm{eos}\rangle\langle/\mathrm{loss}\rangle\) \\
\(\mathcal{D}\oplus\mathcal{P}_{\mathrm{tts}}\) &
\([\mathrm{desc}]\,\mathbf{d}\oplus\mathcal{P}_{\mathrm{tts}}\) \\
\bottomrule
\end{tabular}
}
\caption{Symbolic task templates used in downstream training. \(\mathbf{t}\), \(\mathbf{a}\), and \(\mathbf{d}\) denote text, audio latent sequence, and description instruction, respectively.}
\label{tab:downstream_templates}
\end{table}

\section{Implicit Fidelity Transfer Formulation}
\label{app:proof_fidelity_transfer}

\noindent\textit{\textbf{Proposition 1.}
Let}
\(
\epsilon_{\mathrm{AE}}
=
\mathbb{E}_{\mathbf{x}}
[
\|
\mathbf{x}
-
G_{\psi}(\mathbf{z}_{\mathrm{AE}})
\|_2^2
]
\)
\textit{denote the waveform reconstruction distortion of the Stage-I autoencoder. Assume that the frozen decoder \(G_{\psi}\) is locally \(L_{\psi}\)-Lipschitz in a neighborhood containing both \(\mathbf{z}_{\mathrm{AE}}\) and the variational samples \(\mathbf{z}_{\mathrm{VAE}}\). Define the AE-to-VAE latent shift as:}
\begin{equation}
\delta_{\mathrm{shift}}
=
\mathbb{E}_{\mathbf{x}}
\mathbb{E}_{\mathbf{z}_{\mathrm{VAE}}\sim q_{\eta}(\cdot|\mathbf{z}_{\mathrm{AE}})}
\left[
\left\|
\mathbf{z}_{\mathrm{VAE}}
-
\mathbf{z}_{\mathrm{AE}}
\right\|_2^2
\right].
\end{equation}
\textit{Then the expected waveform distortion of the variational latent satisfies}
\begin{equation}
\begin{aligned}
&\mathbb{E}_{\mathbf{x}}
\mathbb{E}_{\mathbf{z}_{\mathrm{VAE}}\sim q_{\eta}(\cdot|\mathbf{z}_{\mathrm{AE}})}
\left[
\left\|
\mathbf{x}
-
G_{\psi}(\mathbf{z}_{\mathrm{VAE}})
\right\|_2^2
\right]
\\
&\qquad\le
2\epsilon_{\mathrm{AE}}
+
2L_{\psi}^{2}\delta_{\mathrm{shift}} .
\end{aligned}
\end{equation}

\noindent\textbf{Proof.}
For compactness, denote
\(
\hat{\mathbf{x}}_{\mathrm{AE}}=G_{\psi}(\mathbf{z}_{\mathrm{AE}})
\)
and
\(
\hat{\mathbf{x}}_{\mathrm{VAE}}=G_{\psi}(\mathbf{z}_{\mathrm{VAE}})
\).
For any input \(\mathbf{x}\), by adding and subtracting \(\hat{\mathbf{x}}_{\mathrm{AE}}\), we have
\begin{equation}
\begin{split}
\mathbf{x}-\hat{\mathbf{x}}_{\mathrm{VAE}}
&=
\mathbf{x}-\hat{\mathbf{x}}_{\mathrm{AE}}
+
\hat{\mathbf{x}}_{\mathrm{AE}}
-
\hat{\mathbf{x}}_{\mathrm{VAE}}.
\end{split}
\end{equation}
Using \(\|\mathbf{a}+\mathbf{b}\|_2^2 \le 2\|\mathbf{a}\|_2^2+2\|\mathbf{b}\|_2^2\), we obtain
\begin{equation}
\begin{split}
\left\|
\mathbf{x}-\hat{\mathbf{x}}_{\mathrm{VAE}}
\right\|_2^2
&\le
2
\left\|
\mathbf{x}-\hat{\mathbf{x}}_{\mathrm{AE}}
\right\|_2^2 \\
&
\quad+
2
\left\|
\hat{\mathbf{x}}_{\mathrm{AE}}
-
\hat{\mathbf{x}}_{\mathrm{VAE}}
\right\|_2^2.
\end{split}
\end{equation}
By the local \(L_{\psi}\)-Lipschitz continuity of \(G_{\psi}\),
\begin{equation}
\begin{split}
\left\|
\hat{\mathbf{x}}_{\mathrm{AE}}
-
\hat{\mathbf{x}}_{\mathrm{VAE}}
\right\|_2^2
&\le
L_{\psi}^{2}
\left\|
\mathbf{z}_{\mathrm{AE}}
-
\mathbf{z}_{\mathrm{VAE}}
\right\|_2^2.
\end{split}
\end{equation}
Taking expectation over \(\mathbf{x}\) and \(\mathbf{z}_{\mathrm{VAE}}\sim q_{\eta}(\cdot|\mathbf{z}_{\mathrm{AE}})\) gives
\begin{equation}
\begin{split}
\mathbb{E}_{\mathbf{x}}
\mathbb{E}_{q_{\eta}}
\left[
\left\|
\mathbf{x}
-
G_{\psi}(\mathbf{z}_{\mathrm{VAE}})
\right\|_2^2
\right]
\le
2\epsilon_{\mathrm{AE}}
+
2L_{\psi}^{2}\delta_{\mathrm{shift}}.
\end{split}
\tag*{$\square$}
\end{equation}

\section{Experimental Setting and Responsible Use Details}
\label{app:setting_details}

\paragraph{Tokenizer configuration and training settings.}
Table~\ref{tab:tokenizer_config} reports the tokenizer-side parameters. And Table~\ref{tab:tokenizer_training_config} summarizes the optimizer, scheduler, and loss weights used for tokenizer training. All audio is resampled to 48~kHz. The generator is trained with a multi-period discriminator and a multi-scale sub-band CQT discriminator, following the BigVGAN V2 configuration. In Stages I--II, training uses 9.6-second cropped audio segments; in Stage III, the per-GPU batch size is set to 1 to support downstream supervision.

\begin{table}[t]
\centering
\small
\setlength{\tabcolsep}{4pt}
\renewcommand{\arraystretch}{1.08}
\resizebox{\columnwidth}{!}{%
\begin{tabular}{@{}lcc@{}}
\toprule
\textbf{Component} & \textbf{\ourmethod{}-Base} & \textbf{\ourmethod{}-Unite} \\
\midrule
Encoder & 36M & 36M \\
Decoder & 128M & 128M \\
Variational bottleneck & 17M & 17M \\
Semantic encoder & -- & 680M \\
\midrule
Total & 181M & 861M \\
\bottomrule
\end{tabular}
}
\caption{Parameter counts of tokenizer-side representation modules used in downstream modeling.}
\label{tab:tokenizer_config}
\end{table}

\begin{table}[t]
\centering
\small
\setlength{\tabcolsep}{4pt}
\renewcommand{\arraystretch}{1.08}
\resizebox{\columnwidth}{!}{%
\begin{tabular}{@{}ll@{}}
\toprule
\textbf{Setting} & \textbf{Value} \\
\midrule
\multicolumn{2}{@{}l}{\textbf{Optimizer}} \\
Optimizer & AdamW \\
Initial learning rate & \(1\times10^{-4}\) \\
Betas & \((0.8, 0.99)\) \\
Epsilon & \(1\times10^{-6}\) \\
Gradient clipping & 500 \\
\midrule
\multicolumn{2}{@{}l}{\textbf{Scheduler}} \\
Scheduler & exponential decay \\
Final learning rate floor & \(1\times10^{-6}\) \\
Warmup steps & 1 \\
Decay rate & 0.9999996 \\
\midrule
\multicolumn{2}{@{}l}{\textbf{Loss weights}} \\
Adversarial generator loss & 1.0 \\
Feature matching loss & 2.0 \\
Multi-scale mel loss & 45.0 \\
KL loss, Stage II & \(\beta_{\mathrm{low}}=0.1\) \\
KL loss, Stage III & \(\beta_{\mathrm{high}}=7.0\) \\
WavLM distillation loss & 1.0 \\
X-vector distillation loss & 1.0 \\
Supervision CE loss & 1.0 \\
\bottomrule
\end{tabular}
}
\caption{Tokenizer optimization settings and loss weights. HoliTok-Base and HoliTok-Unite use the same recipe, except that the supervision encoder is non-causal for HoliTok-Base and causal for HoliTok-Unite.}
\label{tab:tokenizer_training_config}
\end{table}

\paragraph{Downstream configuration and training settings.}
Table~\ref{tab:downstream_arch_config} summarizes the downstream AR+DiT configuration. \ourmethod{}-Base maps VAE latent patches to the LLM hidden space using an 8-layer PatchEncoder. \ourmethod{}-Unite mean-pools semantic features over each patch and uses a lightweight linear projection before the shared LLM backbone and DiT predictor.
For downstream AR+DiT training, all settings use AdamW with learning rate \(1\times10^{-4}\), betas \((0.9,0.99)\), \(\epsilon=1\times10^{-6}\), bf16 precision, and gradient clipping of 2. The learning rate follows a cosine scheduler with 5000 warmup batches and a minimum learning rate of \(1\times10^{-5}\). The TTS-only setting uses \(\mathcal{P}_{\mathrm{tts}}\), controllable TTS prepends a description instruction as \(\mathcal{D}\oplus\mathcal{P}_{\mathrm{tts}}\), and unified ASR--TTS uses \(\mathcal{P}_{\mathrm{tts}}\) for generation and \(\mathcal{P}_{\mathrm{asr}}\) for recognition. The symbolic templates are defined in Table~\ref{tab:downstream_templates}.

\paragraph{}

\paragraph{Potential risks.}
Because \ourmethod{} supports high-quality speech generation, it may be misused for voice impersonation, spoofing, or misleading synthetic speech if deployed without safeguards. The intended use of the released artifacts is research on speech tokenization and unified spoken language modeling. Practical deployments should include consent-aware data policies, provenance or watermarking mechanisms for generated audio when appropriate, and restrictions against impersonation or deceptive use.

\paragraph{Scientific artifacts, licenses, and intended use.}
This work uses public speech and audio datasets, pretrained model components, baseline tokenizers, and evaluation tools as scientific artifacts\footnote{\href{https://huggingface.co/emotion2vec/emotion2vec_plus_large}{emotion2vec checkpoint.}}\footnote{\href{https://drive.google.com/file/d/1-aE1NfzpRCLxA4GUxX9ITI3F9LlbtEGP/view}{speaker embedding checkpoint.}}\footnote{\href{https://huggingface.co/yfyeung/CLSP}{CLSP checkpoint.}}.
 and artifact creators are cited in Section~4, and the training data mixture and descriptive statistics are summarized there. Third-party datasets and models should be used according to their original licenses and terms of use; internally collected corpora are used only for training and are not redistributed. The released code and checkpoints are intended for research use and will include documentation describing model usage, expected inputs and outputs.

\paragraph{On subjective evaluation.}
We add a preliminary NMOS test with 20 listeners. We use 90 candidates: 30 clean LibriSpeech test-other samples, 30 mixed with AudioSet sounds, and 30 mixed with MUSDB music. Each listener rates 15 random samples (5/5/5). The higher NMOS score means the better reconstruction. ``CR'' means the compression ratio.

\begin{center}
\small
\setlength{\tabcolsep}{4pt}
\begin{tabular}{@{}lcccc@{}}
\toprule
\textbf{Model} & \textbf{CR} & \shortstack{\textbf{Clean}\\\textbf{NMOS}} & \shortstack{\textbf{Complex}\\\textbf{NMOS}} & \shortstack{\textbf{Overall}\\\textbf{NMOS}} \\
\midrule
HoliTok-Base & 7.5x & 4.67 & \textbf{4.23} & \textbf{4.38} \\
HoliTok-Unite & 7.5x & \textbf{4.72} & 4.19 & 4.36 \\
MingTok-Audio & 2.19x & 4.67 & 4.12 & 4.31 \\
Semantic-VAE & 2.73x & 4.41 & 3.14 & 3.57 \\
\bottomrule
\end{tabular}
\end{center}

HoliTok keeps strong perceived quality despite higher compression. It is close to MingTok-Audio on clean speech and better on complex speech, with a compact 25 Hz latent sequence.

\section{Complete Ablation Results}
\label{app:unified_ablation}

\begin{table*}[!t]
\centering
\small
\setlength{\tabcolsep}{3.5pt}
\renewcommand{\arraystretch}{1.08}
\resizebox{\textwidth}{!}{%
\begin{tabular}{@{}lcccccccccc@{}}
\toprule
\multirow{3}{*}{\textbf{Model}}
& \multirow{3}{*}{\textbf{Ablation setting}}
& \multicolumn{6}{c}{\textbf{TTS}}
& \multicolumn{3}{c}{\textbf{ASR}} \\
\cmidrule(lr){3-8}\cmidrule(l){9-11}
&
& \multicolumn{2}{c}{\textbf{Seed-TTS-en}}
& \multicolumn{2}{c}{\textbf{Seed-TTS-zh}}
& \multicolumn{2}{c}{\textbf{Seed-TTS-hard}}
& \textbf{test-clean}
& \textbf{test-other}
& \textbf{AISHELL-1} \\
\cmidrule(lr){3-4}\cmidrule(lr){5-6}\cmidrule(lr){7-8}\cmidrule(l){9-11}
&
& \textbf{WER} $\downarrow$
& \textbf{SIM} $\uparrow$
& \textbf{WER} $\downarrow$
& \textbf{SIM} $\uparrow$
& \textbf{WER} $\downarrow$
& \textbf{SIM} $\uparrow$
& \textbf{WER} $\downarrow$
& \textbf{WER} $\downarrow$
& \textbf{WER} $\downarrow$ \\
\midrule

\multirow{2}{*}{SemanticVAE}
& default
& 102.32 & 0.47
& 99.30  & 0.61
& 97.31  & 0.61
& 9.69 & 21.32 & 15.81 \\
& DiT init
& 34.12 & 0.60
& 4.10 & 0.70
& 29.63 & 0.66
& 10.41 & 25.52 & 16.58 \\

\addlinespace[1pt]
\midrule

\multirow{2}{*}{MingTok-Audio}
& default
& 51.06 & 0.42
& 18.17 & 0.61
& 50.35 & 0.55
& 4.62 & 9.06 & 5.01 \\
& DiT init
& 14.53 & 0.58
& 3.14 & 0.69
& 28.21 & 0.64
& 4.32 & 10.41 & 5.25 \\

\addlinespace[1pt]
\midrule

\multirow{5}{*}{\ourmethod{}-Base}
& default
& 27.85 & 0.52
& 4.40 & 0.66
& 30.44 & 0.61
& 6.45 & 16.51 & 14.92 \\
& DiT init
& 6.63 & 0.59
& 1.71 & 0.69
& 16.97 & 0.66
& 6.39 & 15.92 & 14.32 \\
& w/o distill
& 22.80 & 0.51
& 3.65 & 0.65
& 19.89 & 0.61
& 6.27 & 16.04 & 15.23 \\
& w/o supervise
& 110.05 & 0.39
& 97.56 & 0.60
& 96.81 & 0.60
& 6.44 & 16.59 & 16.63 \\
& w/o. both
& 103.14 & 0.37
& 98.11 & 0.58
& 97.24 & 0.57
& 6.12 & 16.00 & 14.99 \\

\addlinespace[1pt]
\midrule

\multirow{3}{*}{\ourmethod{}-Unite}
& default
& 7.20 & 0.55
& 1.78 & 0.67
& 16.79 & 0.62
& 5.48 & 12.65 & 5.93 \\
& DiT init
& 3.64 & 0.59
& 1.13 & 0.69
& 12.53 & 0.65
& 5.32 & 13.12 & 7.17 \\
& freeze semantic encoder
& 8.46 & 0.56
& 1.86 & 0.66
& 15.60 & 0.63
& 6.44 & 15.92 & 7.42 \\

\bottomrule
\end{tabular}%
}
\caption{Complete ablation results for unified spoken language modeling. TTS is evaluated by WER and SIM on Seed-TTS subsets, and ASR is evaluated by WER on LibriSpeech test-clean/test-other and AISHELL-1. ``default'' denotes the standard unified training setting for each representation. ``DiT init'' initializes the DiT predictor from a TTS-specialized checkpoint. For \ourmethod{}-Base, ``w/o distill'', ``w/o supervise'', and ``w/o both'' remove representation distillation, multi-task supervision, and both objectives in Stage III, respectively. For \ourmethod{}-Unite, ``freeze semantic encoder'' keeps the causal semantic encoder fixed during downstream training.}
\label{tab:unified_slm_ablation}
\end{table*}

Table~\ref{tab:unified_slm_ablation} provides a detailed ablation of the unified spoken language modeling setting. The DiT initialization rows show that a TTS-oriented initialization substantially improves generation across all representations, especially for the two baseline tokenizers whose default unified training yields high TTS WER. This confirms that the downstream DiT head is a major factor for continuous-latent speech generation, but it does not by itself guarantee balanced understanding performance: for example, the initialized \ourmethod{}-Unite improves TTS WER but degrades AISHELL-1 ASR compared with its default setting.

For \ourmethod{}-Base, removing distillation improves several TTS WER scores but slightly weakens speaker similarity and does not improve ASR consistently, suggesting that representation distillation mainly contributes semantic and paralinguistic information rather than pure generation ease. Removing supervision severely degrades TTS while increasing AISHELL-1 WER, indicating that downstream supervision is important for both generation robustness and cross-lingual understanding ability. Removing both distillation and supervision similarly weakens TTS performance, further confirming that downstream-aware enrichment is necessary for a holistic tokenizer. For \ourmethod{}-Unite, freezing the semantic encoder weakens ASR on all three test sets and gives mixed TTS changes, showing that adapting the semantic interface during unified training is important for balancing generation and recognition. Overall, the ablations indicate that strong unified modeling requires both a learnable continuous latent space and task-aware adaptation of the semantic and DiT components.

\FloatBarrier

\section{Use of AI Assistants}
\label{app:ai_assistants}

AI assistants are used to support writing and editing tasks, including grammar checking, wording refinement, and LaTeX formatting. The authors review and edit all AI-assisted text and retain responsibility for the scientific claims, experimental design, analysis, and final manuscript.